\documentclass[12pt,a4paper]{article}
\pdfoutput=1

\usepackage{epsfig}
\usepackage{amsmath}
\usepackage{amssymb}
\usepackage{verbatim}
\usepackage{bm}
\usepackage{dsfont}
\usepackage[footnotesize]{caption}
\usepackage{subfigure}
\usepackage{cancel}
\usepackage{cite}
\usepackage{a4wide}
\usepackage{color}

\usepackage{multirow}

\def\Mp{{M_{\rm P}}}

\def\BL{$B$$-$$L$ }
\def\vp{\varphi}

\usepackage{geometry}
\begin{document}

\begin{minipage}{0.4\textwidth}
 \begin{flushleft}
DESY 21-021
\end{flushleft}
\end{minipage} \hfill
\begin{minipage}{0.4\textwidth}
\begin{flushright}
 February 2021
\end{flushright}
 \end{minipage}

\vskip 1.5cm

\begin{center}
  {\Large\bf Metastable strings and dumbbells\\\vspace{0.2cm}
    in supersymmetric hybrid inflation}

\vskip 1cm

{\large Wilfried~Buchm\"uller
}\\[3mm]
{\it{Deutsches Elektronen-Synchrotron DESY, 22607 Hamburg, Germany}
}
\end{center}

\vskip 1cm

\begin{abstract}
\noindent 
We study symmetry breaking and 
topological defects in a supersymmetric model with gauge
group $\text{U}(2)$, which can be identified with the right-handed
part $\text{SU}(2)_R \times \text{U}(1)_{B-L}$ of an extended electroweak
symmetry of the Standard Model.
The model has two phases of hybrid inflation terminated by tachyonic
preheating where either monopoles and strings or, alternatively,
dumbbells are formed. In the first case a stochastic gravitational wave
background is predicted in the LIGO-Virgo band, possibly extending
to the LISA frequency band and to
nanohertz frequencies, which is generated by a metastable cosmic string
network. In the second case no topological defects survive inflation
and no stochastic gravitational wave background is produced.
\end{abstract}

\thispagestyle{empty}

\newpage

\section{Introduction}

The formation of topological defects is a generic feature of
cosmological phase transitions \cite{Kibble:1976sj}. While monopoles
and domain walls would overclose the universe and must therefore be avoided, cosmic
strings evolve towards a scaling regime where their fraction of the
total energy density is constant. Cosmic strings
have characteristic signatures in gravitational
lensing, the cosmic microwave background  and the stochastic
gravitational wave background (SGWB) and are therefore a potentially very interesting
messenger from the early universe (for reviews and references, see, for
example, \cite{Hindmarsh:2011qj,Auclair:2019wcv}).

Symmetry breaking in grand unified theories (GUTs) can produce many kinds of
topological or non-topological defects, including Nielsen-Olesen
strings \cite{Nielsen:1973cs}, 't Hooft-Polyakov monopoles
\cite{tHooft:1974kcl,Polyakov:1974ek}, unstable ``dumbbells'' or
``X-strings'' connecting a monopole-antimonopole pair 
\cite{Nambu:1977ag},
or ``necklaces'' where more than one string is attached to a monopole 
\cite{Kibble:1982ae}. The monopole problem can be solved by inflation
which is naturally connected to the GUT scale in supersymmetric hybrid
inflation \cite{Copeland:1994vg,Dvali:1994ms}. Moreover, it has been
shown for a large class of GUTs that in spontaneous symmetry breaking
schemes solving the monopole problem cosmic string formation is
unavoidable \cite{Jeannerot:2003qv}.
 
The seesaw mechanism \cite{Minkowski:1977sc,Yanagida:1979as,GellMann:1980vs}
connects the GUT scale with neutrino masses, and the related
leptogenesis scenario \cite{Fukugita:1986hr} provides an elegant explanation of the
cosmological baryon asymmetry (for a recent review, see \cite{Bodeker:2020ghk}).
The masses of the heavy seesaw partners of the light neutrinos depend on the breaking
scale of \BL, the difference of baryon and lepton number. In a
detailed study \cite{Buchmuller:2012wn,Buchmuller:2014epa} it was
shown that cosmological \BL breaking,
starting from a false vacuum of unbroken \BL, can lead to a consistent
picture of hybrid inflation, leptogenesis and dark matter. The
spontaneous breaking of $\text{U}(1)_{B-L}$ is tied to the formation of a cosmic
string network and the corresponding SGWB was evaluated in \cite{Buchmuller:2013lra}.

Recently, it was pointed out that the seesaw mechanism and
leptogenesis can be tested by means of metastable cosmic strings
\cite{Dror:2019syi} that evade the tension between GUT-scale
strings and the upper bound on the string tension from pulsar timing
array (PTA) experiments \cite{Arzoumanian:2018saf,Kerr:2020qdo,Shannon:2015ect}.
Embedding the $\text{U}(1)_{B-L}$ model into an $\text{SO}(10)$ GUT,
the \BL strings become metastable \cite{Buchmuller:2019gfy}. They
decay by quantum tunneling into string segments connecting
monopole-antimonopole pairs. In the semiclassical
approximation the decay rate per string unit length is given by
\cite{Vilenkin:1982hm, Preskill:1992ck,Leblond:2009fq,Monin:2008mp}
\begin{equation}\label{decayrate}
  \Gamma_d = \frac{\mu}{2 \pi} \exp\left( - \pi \kappa \right) \ ,\quad
  \kappa = \frac{m_M^2}{\mu} \ ,
\end{equation}
where $m_M$  is the monopole mass and $\mu$ is the string tension. Using
this decay rate and the BOS model \cite{Blanco-Pillado:2013qja} for GW
emission from string loops (for a general discussion, see \cite{Auclair:2019wcv}),
the SGWB generated by a metastable string network was evaluated for the
parameters of the \BL model. In the LIGO-Virgo frequency band a GW
signal close to the current upper limit was predicted\footnote{The
  predicted range of the
  dimensionless string tension is $1.0 \times 10^{-7} \leq G\mu \leq
  5.6 \times 10^{-7}$, with Newton's constant $G = 6.7\times
  10^{-39}~\text{GeV}^{-2}$. This is qualitatively consistent with the
  recent LIGO-Virgo upper bound on $G\mu$ for model A \cite{Abbott:2021ksc}. A
  quantitative comparison is not possible since the calculation in
  \cite{Buchmuller:2019gfy} uses an average power spectrum inferred
  from numerical simulations, assuming ``cusp dominance'', whereas in
  model A the GW spectrum is computed from the superposition of
  individual bursts, assuming ``kink dominance''.}
  and from the PTA
bounds an upper bound on the monopole-string-tension ratio was
obtained, $\sqrt{\kappa} \lesssim 8$ \cite{Buchmuller:2019gfy}.

A few months ago, the NANOGrav collaboration has reported evidence for
a stochastic process at nanohertz frequencies
\cite{Arzoumanian:2020vkk}, which has been interpreted as SGWB in a
large number of recent papers. Possible cosmological interpretations
include stable \cite{Ellis:2020ena,Blasi:2020mfx}  as well as
metastable strings \cite{Buchmuller:2020lbh}. In the latter case an analysis
of the NANOGrav data deduced a
monopole-string-tension ratio in the range $7.8 \leq\sqrt{\kappa} \leq
9.0$ (2$\sigma$ CL). Different string models have been compared in
\cite{Blanco-Pillado:2021ygr}.
The possible connection of the NANOGrav results
with GUT models and high-scale leptogenesis has also been studied in
\cite{Chigusa:2020rks,Chakrabortty:2020otp,Samanta:2020cdk,King:2020hyd},
and effects of low-scale leptogenesis on the SGWB were considered in 
\cite{Blasi:2020wpy}. Note that, independent of grand unification, a
SGWB form a cosmic string network is a very interesting and well
motivated signature of physics beyond the Standard Model
\cite{Cui:2017ufi,Cui:2018rwi,Gouttenoire:2019kij,Gouttenoire:2019rtn}.

The determination of the monopole-string-tension ratio $\sqrt{\kappa}$
from an interpretation of the NANOGrav data gives rise to the question
whether the obtained value can be naturally realized in extensions of
the Standard Model. The purpose of this paper is to clarify this
question. In the following we study the simplest possible extension
of the supersymmetric $\text{U}(1)_{B-L}$ model considered in
\cite{Buchmuller:2012wn}, a supersymmetric model with gauge group
$\text{U}(2)$,  corresponding to $\text{SU}(2)_R \times \text{U}(1)_{B-L}/\mathbb{Z}_2$,
which allows for monopoles, strings and dumbbells as topological
defects.

The supersymmetric $\text{U}(2)$ model has three Higgs fields, one
$\text{SU}(2)$ triplet and two $\text{SU}(2)$ doublets.
In Section~2 we discuss the possible symmetry breakings 
and the associated defects, and in Section~3 we analyze the
implications for hybrid inflation. As we shall see, there are always
two phases of inflation which are terminated by tachyonic preheating
\cite{Felder:2000hj},
except for one case where the symmetry is already completely broken
after the first phase of inflation. The sequence of symmetry breakings
is determined by the masses of the triplet and doublet Higgs fields.
Depending on their ratio, either monopoles and metastable strings
or, alternatively, dumbbells occur as topological defects.
In the first case, after inflation a metastable
cosmic string network is formed that leads to a SGWB in the
LIGO-Virgo band, possibly extending to nanohertz frequencies.
However, it is also possible that no cosmological defects survive inflation
and no SGWB is generated. In the first case the reheating mechanism of the
$\text{U}(1)_{B-L}$ model is realized, including leptogenesis and dark
matter production. In the second case, this reheating mechanism is lost.
We conclude in Section~4.

\section{Symmetry breaking in a supersymmetric U(2) model}

The starting point of cosmological $B-L$ breaking is a supersymmetric
Abelian Higgs model with two chiral superfields\footnote{We follow the
conventions of Ref.~\cite{Wess:1992cp}.} $S$ and $S_c$,
carrying charge $q$ and $-q$, respectively, and a gauge singlet
$\phi$. Kahler potential and superpotential are given by
\begin{equation}\label{KPU1}
K = S^\dagger e^{2gqV} S + S_c^\dagger e^{-2gqV}S_c +
\phi^\dagger\phi\ ,\quad
P = \frac{1}{4} W W + \lambda\phi\left(\frac{v_s^2}{2} - SS_c\right)\ ,
\end{equation}
where $V$ is a vector superfield, $W$ is the
supersymmetric field strength, and $v_s$ is chosen real and positive.
The scalar potential is given in terms of the auxiliary fields of vector and
chiral superfields,
\begin{equation}
\mathcal{V} = \frac{1}{2} D^2 + |F_S|^2 + |F_{S_c}|^2 + |F_\phi|^2 \ ,
\end{equation}
which are determined by the equations of motion,
\begin{equation}
  \begin{split}
D &= - gq (|S|^2 - |S_c|^2) \ , \\ 
F_S^* &= \lambda\phi S_c\ , \quad  
F_{S_c}^* = \lambda\phi S\ , \quad
F_\phi^* = -\lambda\left(\frac{v_s^2}{2} - SS_c\right)\ .
\end{split}
\end{equation}
Adding to the scalar potential the kinetic terms for scalar and vector
fields, one obtains for the
bosonic part of the Lagrangian\footnote{We denote the vector component
of $V$ by $A_\mu$, its field strength by $F_{\mu\nu}$, and we use the
same symbol for a chiral superfield and its scalar component.}  
\begin{equation}
\mathcal{L}_b = -\frac{1}{4} F_{\mu\nu}F^{\mu\nu} - (D_\mu S)^*(D^\mu S)
- (D_\mu S_c)^*(D^\mu S_c) - \partial_\mu \phi^* \partial^\mu \phi -
\mathcal{V} \ ,
\end{equation}
with the covariant derivatives $D_\mu S = (\partial_\mu + igq A_\mu)S$ and
$D_\mu S_c = (\partial_\mu - igq A_\mu)S_c$.

The D-term leaves a flat direction, $|S|^2 = |S_c|^2$, and the model has
a continuum of supersymmetric ground states that break the $\text{U}(1)$ symmetry,
\begin{equation}
  S = \frac{v_s}{\sqrt{2}} e^{i\alpha}\ , \quad
  S_c = S^*\ , \quad \phi = 0\ .
\end{equation}
Shifting $S$ and $S_c$ around their vacuum expectation values (vevs),
$S = v_s e^{i\alpha}/\sqrt{2} + S'$, $S_c = v_s e^{-i\alpha}/\sqrt{2} + S'_c$,
the Goldstone multiplet $(S'-S'_c)/\sqrt{2}$ mixes with $V$, yielding a massive
vector multiplet with mass $m_V = \sqrt{2}gqv_s$, and $(S'+S'_c)/\sqrt{2}$
and $\phi$ form a massive chiral multiplet with mass $m_S = \lambda v_s$.

The vacuum manifold $\mathcal{M}$ is a circle $S^1$, as in the
non-supersymmetric case, with non-trivial first homotopy group, 
$\pi_1(\mathcal{M}) = \mathbb{Z}$. Hence, the model has topologically
stable strings as exited states. Along the D-flat direction they are
given by the Nielsen-Olesen string solutions \cite{Nielsen:1973cs}. In cylindrical
coordinates $(\rho,\varphi,z)$ the field configurations of static strings with
winding number $n$ along the $z$ axis read
\begin{equation}
  S = \frac{v_s}{\sqrt{2}}f(\rho)e^{ni\varphi} = S_c^*\ , \quad
  A_0 = 0\ , \quad A_i = -\frac{n}{g\rho} h(\rho) \partial_i\varphi\ ,
\end{equation}
where the functions $f$ and $h$ satisfy the boundary conditions
$f(0) = h(0) =0$,  $f(\infty) = h(\infty) = 1$. The total magnetic
flux along the string is $2n\pi/g$ and the string tension, the energy
per unit length, is given by
$\mu = 2\pi B(\beta) v_s^2 = \pi m_V^2B(\beta)/(gq)^2$, where
$\beta = m_S^2/m_V^2 = \lambda^2/2g^2$ and $B$ is a slowly varying
function with $B(1)=1$ \cite{Hindmarsh:2011qj}. In the following we shall consider Type-I
strings with $\beta < 1$, which are stable and feel an attractive force. 

Cosmic strings formed in a phase transition are a source of gravitational
waves. As described in the introduction, this leads to important
constraints on the allowed string trension. These constraints are
considerably weaker for metastable strings which naturally occur in
GUT-symmetry-breaking phase transitions. Closed strings can then break
into string segments connecting a monopole and an antimonopole. The
simplest case where this occurs is the breaking of
$\text{SU}(2)\times \text{U}(1)$, or more precisely
$\text{U}(2) =\text{SU}(2)\times \text{U}(1)/\mathbb{Z}_2$,
to $\text{U}(1)$. The corresponding  vacuum manifold is
$S^3$, which contains $S^1 \cup S^2$, the union of the vacuum manifolds of
stable strings and monopoles.

Consider first the breaking of $\text{SU}(2)$ to $\text{U}(1)$ by a
$\text{SU}(2)$ triplet $U^a$, $a = 1,..,3$, which leads to topologically
stable 't~Hooft-Polyakov monopoles \cite{tHooft:1974kcl,Polyakov:1974ek}. Kahler potential and
superpotential of the supersymmetric theory read
\begin{equation}\label{KPSU2}
K = U^\dagger e^{2gV} U \ ,\quad
P = \frac{1}{8} \text{tr}[W W] \ ,
\end{equation}
where $V = V^a T^a$ and $(T^a)_{bc} = -i\epsilon_{abc}$. The
bosonic part of the Lagrangian is given by
\begin{equation}
\mathcal{L}_b = -\frac{1}{4} F^a_{\mu\nu}F^{a\mu\nu} - (D_\mu U^a)^*(D^\mu U^a)
- ig \epsilon_{abc} D^a U^{b *} U^c + \frac{1}{2} D^aD^a +
F^{a *}_U F^a_U \ .
\end{equation}
where $(D_\mu U)^a = \partial_\mu U^a - g \epsilon_{abc} A^b_\mu U^c$ and
$F^a_{\mu\nu} = \partial_\mu A^a_\nu - \partial_\nu A^a_\mu
-g\epsilon_{abc}A^b_\mu A^c_{\nu}$. Note that this is the well known
$\text{SU}(2)$ Super-Yang-Mills theory with $\mathcal{N}=2$
supersymmetry\footnote{A supersymmetric extension of the Standard
  Model containing a $\mathcal{N} = 2$ sector with gauge group
  $\text{SU}(2)_R$ could arise in an orbifold compactification of a
  supersymmetric Pati-Salam or $\text{SO}(10)$ theory in five or six
  dimensions, breaking the GUT group to the Standard Model gauge group.}
The classical theory has a flat direction,
$U^a = u \delta_{a3}$ up to gauge transformations.
For this theory the full quantum and nonperturbative corrections are
known. They preserve the flat direction which interpolates between
a confinement phase with monopole condensation and a perturbative
Higgs phase \cite{Seiberg:1994rs}.

We are interested in the Higgs phase of the model, with $u$ much
larger than the confinement scale $\Lambda$. As in the Abelian Higgs
model, the vacuum degeneray can be lifted by adding a term to the
superpotential in Eq.~\eqref{KPSU2},
\begin{equation}\label{PSU2}
  P = \frac{1}{8} \text{tr}[W W] + \frac{\lambda'}{2} \phi' \left(v_u^2 -
    U^T U\right) \ ,
\end{equation}
where $\phi'$ is a gauge singlet superfield and $v_u$ is chosen real and
positive. The additional term in the superpotential breaks $\mathcal{N}=2$
supersymmetry to $\mathcal{N}=1$ supersymmetry.
The equations of motion for the auxiliary fields are 
\begin{equation}
D^a =  ig \epsilon_{abc} U^{b*} U^c \ , \quad
F_U^{a*} = \lambda' \phi' U^a \ , \quad
F_{\phi'}^* = -\frac{\lambda'}{2} \left(v_u^2 - U^T U\right)
\ .
\end{equation}
The model has a supersymmetric vacuum, 
\begin{equation}
U^a = v_u \delta_{a3}\ , \quad \phi' = 0 \ ,
\end{equation}
where $U^a$ is determined up to a $\text{SU}(2)$ rotation. A
$\text{U}(1)$ subgroup of $\text{SU}(2)$ remains unbroken. There is one
massless vector multiplet, $V^3$, and a charged vector
multiplet with mass $m_V = \sqrt{2}gv_u$, which contains the Goldstone
multiplets $U^{1,2}$. The chiral multiplets ${U'}^3=U^3-v_u$ and $\phi'$ form a
massive multiplet with mass $m_U = \lambda' v_u$. 

As in the non-supersymmetric case the vacuum manifold is a 2-sphere $S^2$
with non-trivial homotopy group $\pi_2(\mathcal{M}) = \mathbb{Z}$. Hence,
there are topologically stable monopoles as excited states. The
simplest ``hedgehog'' field configuration is given by \cite{tHooft:1974kcl,Polyakov:1974ek}
\begin{equation}\label{hedgehog}
  U^a = v_u f(r) \frac{x^a}{r}\ , \quad A^a_0 = 0 \ , \quad
  A^a_i = h(r) \epsilon_{aij} \frac{x^j}{gr^2} \ ,
\end{equation}
where $r=(x^i x^i)^{1/2}$; the functions $f$ and $h$ satisfy the boundary conditions
$f(0) = h(0) =0$,  $f(\infty) = h(\infty) = 1$. The scalar field, and
therefore the unbroken symmetry generator, points in radial direction,
$\hat{\phi^a} = x^a/r$, and at large distances the gauge invariant
magnetic field strength is
\begin{equation}
   B_i = -\frac{1}{2}\hat{\phi}^a \epsilon_{ijk} F^a_{jk} = \frac{x^i}{gr^3} \ .
  \end{equation}
 Hence, the magnetic charge of the monopole is $4\pi/g$. The mass
 satisfies the Bogomol'nyi bound \cite{Bogomolny:1975de}
 \begin{equation}
   m_M \geq \frac{4\pi m_V}{g^2}\ ,
 \end{equation}   
which is saturated in the Prasad-Sommerfield limit $\lambda'/g
\rightarrow 0$ \cite{Prasad:1975kr}.

We are interested in embedding $\text{U}(1)_{B-L}$ into a 
larger group such that \BL strings can break into segments
connecting a monopole with an antimonopole. The simplest possibility
is to extend the electroweak part of the standard model gauge group
to $\text{SU}(2)_L\times\text{SU}(2)_R \times \text{U}(1)_{B-L}$ which
is then spontaneously broken to $\text{SU}(2)_L\times \text{U}(1)_Y$,
with $Y = T^3_R + \frac{1}{2}(B-L)$.
For the symmetry breaking 
$\text{SU}(2)_R\times \text{U}(1)_{B-L}$
to $\text{U}(1)_Y$, the vacuum manifold is $\mathcal{M} =
\text{U}(2)/\text{U}(1) = S^3$. Hence, the homotopy groups $\pi_1(\mathcal{M})$
and $\pi_2(\mathcal{M})$ are trivial and there are neither topologically
stable monopoles nor strings. However, as we shall see, there can be
metastable strings or unstable dumbbells.

Let us now consider a supersymmetric $\text{U}(2)$ model with a
triplet $U$ and a pair of
doublets $S$, $S_c$, transforming as $U \sim (3,0)$ and $S \sim
(2,q)$, $S_c \sim (\bar{2},-q)$ with respect to $\text{SU}(2)\times\text{U}(1)$.
For previous discussions of defects in non-supersymmetric
$\text{U}(2)$ models with triplet and  doublet scalars, see, for example,
\cite{Copeland:1987ht,Kephart:1995cg,Achucarro:1999it,Kibble:2015twa}. 
For the supersymmetric model, we choose
Kahler potential and superpotential as a combination of
Eqs.~\eqref{KPU1}, \eqref{KPSU2} and \eqref{PSU2}, with an additional
mass term in the superpotential\footnote{Note that this model is
  different from standard left-right symmetric models where 
  triplet fields carry $\text{U}(1)$ charge and therefore occur in
  pairs.  See, for example, \cite{Aulakh:1998nn,Babu:2008ep}.}
\begin{equation}\label{KPU2}
  \begin{split}
  K &= U^\dagger e^{2gV} U + S^\dagger e^{2(g\tilde{V} + g'qV')}S
  + S_c^\dagger e^{-2(g\tilde{V}+g'qV')}S_c + \phi^\dagger\phi +
  {\phi'}^{\dagger}\phi'\ , \\
  P &= \frac{1}{8} \text{tr}[W W] + \frac{1}{4} W'W' + 2 h S^T_c\tilde{U} S \\
  &\ + \frac{\lambda'}{2} \phi' \left(v_u^2 - U^T U\right)
  + \lambda\phi\left(\frac{v_s^2}{2} - S^T_cS\right) - h v_u S^T_cS\ ,
\end{split}
\end{equation}
where $U=(U^1,..,U^3)^T$, $V = V^a T^a$, $\tilde{V} = V^a \tau^a/2$
and $\tilde{U} = U^a \tau^a/2$;
$V'$ is the $\text{U}(1)$ vector field and $W'$ is the supersymmetric field strength.
The term cubic in the Higgs fields couples the fundamental triplet
$U^a$ to the composite triplet $S^T_c\tau^a S$. Without this term the
group $\text{U}(2)$ would in general be completely broken, and no
$\text{U}(1)$ subgroup would survive. The mass
term $S^T_cS$ is needed in order to obtain a supersymmetric vacuum with
$\langle P\rangle = 0$. Supersymmetry breaking, induced by another sector,
can then be hierarchically smaller than a possible \BL contribution
to the gravitino mass, given by $\langle P\rangle/M^2_{\text{P}}  \sim hv_uv_s^2/M^2_{\text{P}}$.

The equations of motion for the auxiliary fields are
\begin{equation}
  \begin{split}
 D^a &=  -g\left( -i\epsilon_{abc} U^{b*} U^c +
 \frac{1}{2}S^\dagger \tau^a S -  \frac{1}{2}S_c^\dagger \tau^a S_c\right)\ ,\\ 
 D &= -g'q (S^\dagger S - S_c^\dagger S_c) \ , \\
F_U^{a*} &= - h S^T_c\tau^a S  + \lambda' \phi' U^a \ , \\
F_{\phi'}^* &= -\frac{\lambda'}{2} \left(v_u^2 - U^T U\right) \ ,\\
F_S^\dagger &=  -S^T_c (2h \tilde{U} - \lambda\phi  - hv_u) \ , \\   
F_{S_c}^* &=   -(2h \tilde{U}  - \lambda\phi - hv_u)S \ , \\
F_\phi^* &= -\lambda\left(\frac{v_s^2}{2} - S^T_cS\right)\ .
\end{split}
\end{equation}
The D-terms have again flat directions and
a supersymmetric vacuum exists with the expetation values
\begin{equation}
  U^a = v_u \delta_{a3}\ , \quad
   S = S_c = 
  \frac{v_s}{\sqrt{2}}\left(\begin{array}{c} 1 \\
                                               0 \end{array}\right) \
                                           , \quad
\phi' = \frac{hv_s^2}{2\lambda' v_u} \ , \quad
                                           \phi = 0 \ .
                                         \end{equation}
Note that the triplets $U^a$ and $S^T_c\tau^aS$ are parallel, which is
enforced by $h\neq 0$. Otherwise, the relative orientation of the two
triplets would not be fixed.                                         

In order to obtain the mass spectrum of the model, one has to shift
the chiral multiplets around their vacuum expectation values:
$U^3 = v_u + {U'}^3$, $S=(v_s/\sqrt{2} + {S^0}',S^-)^T$, $S_c=(v_s/\sqrt{2} +
{S^0_c}',S^+)^T$, $\phi' = hv_s^2/(2\lambda'v_u) + \hat{\phi}$. From
the terms of the Kahler potential \eqref{KPU2} that are linear in $V$,
$\tilde{V}$ and $V'$, one obtains the Goldstone
multiplets $\Pi^\mp= (2v_u U^\mp + v_s S^\mp)/\sqrt{4v_u^2 + v_s^2}$
and $\Pi^0 = ({S^0}'-{S^0_c}')/\sqrt{2}$, where
$U^\pm = (U^1 \mp i U^2)/\sqrt{2}$. They are absorbed by the vector
multiplets $V^\pm = (V^1 \mp iV^2)/\sqrt{2}$ and $V_X = (g V^3 + 2g'q
V')/(2g_X)$, respectively. The orthogonal vector multiplet
$V_Y = (-2g'q V^3 + gV')/(2g_X)$ remains massless. The masses
of the vector multiplets are
\begin{equation}\label{mVmX}
  m^2_V = g^2 \left(2 v_u^2 + \frac{v_s^2}{2}\right) \ , \quad
  m^2_X = 2g_X^2 v_s^2  \ ,
\end{equation}
with 
\begin{equation}  
  g_X = \left({g'}^2q^2 + \frac{g^2}{4}\right)^{1/2}\ , \quad
  \tan{\Theta} = \frac{2g'q}{g} \ ,
  \end{equation}
where the angle $\Theta$ is the analogue of the weak angle in the
electroweak theory.
  
The mass matrix of the remaining six chiral multiplets is
given by the quadratic part of the superpotential,
\begin{equation}
  P_m = - h\left(\frac{4v_u^2+v_s^2}{v_u}\right) \Sigma^-\Sigma^+ 
- \lambda' v_u \hat{\phi} {U'}^3 - v_s(\lambda \phi - h{U'}^3)\Sigma^0
  - \frac{h v_s^2}{4 v_u} ({{U'}^3})^2 \ ,
\end{equation}
where $\Sigma^\mp = (-v_s U^\mp + 2v_u S^\mp)/\sqrt{4v_u^2 + v_s^2}$
and $\Sigma^0 = ({S^0}' + {S^0_c}')/\sqrt{2}$ are orthogonal to the
Goldstone multiples $\Pi^\mp$ and $\Pi^0$, respectively. For $h=0$, $P_m$ is the
sum of the mass terms obtained above for independent $\text{SU}(2)$
and $\text{U}(1)$ breakings.

Depending on the parameters, the model can have two subsequent phase
transitions where first $\text{SU}(2)$ is broken by
$\langle U^a\rangle$ to $\text{U}(1)$, corresponding to $\text{SU}(2)_R
\rightarrow \text{U}(1)_R$, and in a
second step $\text{U}(1)\times \text{U}(1)$ is further broken by
$\langle S\rangle$ and $\langle S_c\rangle$ to a $\text{U}(1)$
subgroup, corresponding to $\text{U}(1)_R \times \text{U}(1)_{B-L}
\rightarrow \text{U}(1)_Y$. This is analogous to the Standard Model,
where $\text{SU}(2)_L$ contains the subgroup $\text{U}(1)_L$
and $\text{U}(1)_L \times \text{U}(1)_Y$ is broken to the
electromagnetic subgroup $\text{U}(1)_Q$.
The topological defects formed in the two phase transitions are monopoles and strings,
respectively. Monopole mass and string tension are approximately
given by 
\begin{equation}\label{monotension}
  m_M \simeq \frac{4\pi m_V}{g^2}\ , \quad \mu \simeq \frac{\pi
    m^2_X}{g_X^2} \ ,
\end{equation}
with $m_V$ and $m_X$ given in Eq.~\eqref{mVmX}.
Note that triplet and doublet vev's both contribute to the monopole mass.

In Pati-Salam- or $\text{SO}(10)$-GUT extensions of the Standard Model
the doublets $S$ and $S_c$ are embedded into Pati-Salam $(4,1,2)$-
and $(\bar{4},1,\bar{2})$-plets, or into $16$- and $\bar{16}$-plets
of $\text{SO}(10)$, respectively\footnote{Heavy Majorana neutrino masses can
  be generated by the nonrenormalizable operator
  $1/M_* h_{ij} S^T L^c_i S^T L^c_j$ where $L^c_i = (n^c_i,e^c_i)^T$,
  $i=1,..,3$, denote the $\text{SU}(2)_R$ doublets of right-handed neutral and charged
  leptons and $h_{ij}$ are Yukawa couplings.}.
The normalization condition
$q^2\text{tr}[(B\!-\!L)^2] = \text{tr}[(T^3_R)^2]$ implies
$q=\sqrt{3/8}$ which, together with
gauge coupling unification, $g=g'$, yields $\tan{\Theta}=\sqrt{3/2}$.
For the parameter $\kappa$ that controls the metastability of
cosmic strings, one obtains
\begin{equation}\label{kappa}
  \kappa = \frac{m_M^2}{\mu} \simeq \frac{ 4 \pi}{g^2 \cos^2{\Theta}}
  \left(\frac{m_V}{m_X}\right)^2\ .
\end{equation}

As discussed in the
introduction, an explanation of the NANOGrav results in terms of
metastable cosmic strings requires $\sqrt{\kappa} \simeq 8$.
At the GUT scale, one has $g^2 \simeq 1/2$. With $\cos^2{\Theta}
= 2/5$, one obtains $\kappa \simeq 20 \pi (m_V/m_X)^2$.
Hence, for vector-boson masses of similar size, i.e.~$m_V \simeq m_X$,
$\sqrt{\kappa} \simeq 8$ results purely from geometrical
factors and the value of the gauge coupling at the GUT scale.

An important open question concerns the range of validity of the
relation \eqref{kappa}. For $m_V \simeq m_X$, the initial separation of the
monopole-antimonopole pair is $L = 2m_M/\mu = 2/\cos^2{\Theta}~m_X^{-1}
= 5~m_X^{-1}$. Hence, the thin-defect approximation 
made in the derivation of the decay rate \eqref{decayrate},
$L \gg m_X^{-1}$,
is only marginally satisfied. Moreover, the transition from metastable
strings to unstable X-strings clearly requires a deeper analysis.

Alternatively, $\text{U}(2)$ can be broken to $\text{U}(1)$ by vev's
of the doublets $S$ and $S_c$ only. The condition for this to happen
will be discussed in the following section.
In this case dumbbells or
X-strings form which are completely analogous to the Z-strings
of the Standard Model \cite{Nambu:1977ag}.
For $\tan{\Theta} = \sqrt{3/2}$,
X-strings are known to be unstable \cite{Vachaspati:1992fi,Achucarro:1999it}.

\section{Hybrid inflation and the formation of monopoles, strings and dumbbells in tachyonic preheating}

A sequence of cosmological phase transitions can lead to different
types of topological defects. In our case there are two phase
transitions, characterized by the vacuum expectation values $v_s$ and
$v_u$. For a thermal history, the sequence of the transitions is not
determined by the symmetry breaking scales but rather by the
corresponding Higgs masses, i.e., $\lambda v_s$ and $\lambda' v_u$
\cite{Kibble:2015twa}. As we shall see, the same is true for the different phases of
hybrid inflation.

Consider the evolution of the system of scalar fields along D-flat directions 
where the relevant part of the scalar potential is given by the F-term contributions,
\begin{equation}\label{Fpotential}
\begin{split}
\mathcal{V}_F &= (h S^T_c\tau^a S  - \lambda' \phi' U^a)^*
  (h S^T_c\tau^a S  - \lambda' \phi' U^a)
+\frac{{\lambda'}^2}{4} |v_u^2 - U^TU|^2 \\
&+S_c^T (2h \tilde{U} - \lambda\phi  - hv_u)
(2h \tilde{U} - \lambda\phi  - hv_u)^\dagger S_c^* \\   
&+ S^\dagger(2h \tilde{U}  - \lambda\phi - hv_u )^\dagger
(2h \tilde{U}  - \lambda\phi - hv_u )S  + \frac{\lambda^2}{4} |v_s^2 - 2 S_c^TS|^2\ .
\end{split}
\end{equation}
The gauge-singlet fields $\phi$ and $\phi'$ can play the role of
inflatons. For $|\phi|, |\phi'| \gg v_s, v_u$ the potential becomes
\begin{equation}
  \mathcal{V}_F \simeq \frac{{\lambda'}^2}{4} v_u^4 +
  \frac{\lambda^2}{4} v_s^4 + {\lambda'}^2 |\phi'|^2 U^\dagger U
+ \lambda^2 |\phi|^2 (S^\dagger S + S_c^\dagger S_c)\ .
\end{equation}
The large mass terms for $U$, $S$ and $S_c$ force the waterfall fields to
zero, the vacuum energy drives inflation and the potential is flat in
$\phi'$ and $\phi$. For simplicity, we choose the inflaton fields
real, i.e., $\phi \equiv \vp/\sqrt{2} = \phi^*$ and
$\phi' \equiv \vp'/\sqrt{2} = {\phi'}^*$.
A small slope of the scalar potential is generated by the one-loop quantum
corrections \cite{Dvali:1994ms}
\begin{equation}
  \mathcal{V}_{1l} \simeq \frac{3 m_U^4}{128\pi^2}
  \ln{\left(\frac{{\vp'}^2}{v_u^2}\right)}
+ \frac{m_S^4}{32\pi^2}\ln{\left(\frac{{\vp}^2}{v_s^2}\right)}\ ,
  \end{equation}
  where $m_U = \lambda' v_u$ and $m_S=\lambda v_s$ are the Higgs masses
  associated with the breaking of $\text{SU}(2)$ and $\text{U}(1)$,
  respectively, as discussed in the previous section.
  During the inflationary phase the time evolution of $\vp$ and $\vp'$
is determined by the slow-roll equations of motion
\begin{equation}\label{slowroll}
3 H \dot{\vp}' = -\partial_{\vp'} \mathcal{V}_{1l}\ , \quad
  3 H \dot{\vp} = -\partial_{\vp} \mathcal{V}_{1l}\ ,
  \end{equation}
  where $H$ is the Hubble parameter during inflation. Starting at
  large field values $\vp'_i$ and $\vp_i$ at an initial time $t_i$,
  $\vp'$ and $\vp$ roll towards their critical values $v_u$ and $v_s$, respectively,
  where tachyonic instabilities occur. Which phase transition
  occurs first depends on how fast the two inflatons run,
which is determined by the ratio of Higgs masses $m_S/m_U$.

\subsection{Formation of metastable strings}

Consider first the case $m_U \gg m_S$. Starting at $\vp_i, \vp'_i \gg
v_s,v_u$, one finds from Eq.~\eqref{slowroll}
that $\vp'$ reaches its critical value $v_u$ at the time
\begin{equation}
t'_c = t_i + \frac{32\pi^2H_i}{m_U^4}\left({\vp'_i}^2 - v_u^2\right) \ ,
\end{equation}
 where the Hubble parameter is $H_i = \sqrt{{\lambda'}^2 v_u^4 +
  \lambda^2v_s^4}/(2\sqrt{3}\Mp)$. At this time the inflaton field $\vp$
takes the value
\begin{equation}
  \vp(t'_c)^2 = \vp_i^2 -
 \frac{4}{3} \left(\frac{m_S}{m_U}\right)^4 \left({\vp_i'}^2 - v_u^2\right)
  \simeq \vp_i^2 - \frac{4}{3} \left(\frac{m_S}{m_U}\right)^4 {\vp'_i}^2\ .
\end{equation}
Hence, with $v_u \sim v_s$ and assuming $\vp'_i \sim \vp_i$, a ratio
$\lambda/\lambda' \sim 0.1$ is already enough to ensure that at $t_c'$
the inflaton field $\vp$ is still large, $\vp(t'_c) \sim \vp_i$, and therefore
far away from its critical value $v_s$. For $\vp' \sim v_u$, the
relevant part of the scalar potential reads
\begin{equation}\label{tachyon1}
  \begin{split}
  \mathcal{V}_F &\simeq \frac{\lambda^2}{4} v_s^4 
  + \frac{\lambda^2}{2} \vp^2 \big(S^\dagger S + S_c^\dagger S_c\big)
+ \frac{m_S^4}{32\pi^2}\ln{\left(\frac{\vp^2}{v_s^2}\right)} \\
  &+ \frac{{\lambda'}^2}{4} \left({v'}^4 + 2\big({\vp'}^2- v_u^2\big) u^Tu 
+ 2\big({\vp'}^2+ v_u^2\big) w^Tw
+\big(u^Tu-w^Tw\big)^2 + 4 (u^Tw)^2\right) \ ,
\end{split}
\end{equation}
where we have split $U$ into real and imaginary parts, $U^a = u^a + i w^a$.

The masses of $S$ and $S_c$ are still large and these waterfall fields therefore  
remain frozen at zero. Also the mass squared of $w^a$ is positive so
that the imaginary part of $U^a$ stays at zero. However, the real part
of $U^a$ becomes tachyonic, which leads
to a rapid growth of the low-momentum ($k < k_*$) quantum fluctuations
and hence of the variance $\langle (u^a)^2(t)\rangle$.
This process of ``tachyonic preheating'' has been investigated
numerically in~\cite{Felder:2000hj}, neglecting the Hubble expansion. The root mean square value of the waterfall field can be treated
as a homogeneous background field, $u^a(t) \simeq \langle
(u^a)^2(t)\rangle^{1/2}$ within a patch of the size of the coherence
length $\sim k_*^{-1}$.
The growth of the fluctuations is terminated by backreaction, i.e., by
the self-interaction of the waterfall field, as
the different modes scatter off each other.
The onset of the phase transition has also
been studied analytically taking the Hubble expansion into account 
\cite{Asaka:2001ez} and it has been shown that waterfall field and
inflaton reach their vacuum values within a few Hubble times.
The vacuum value of $u^a$ can be chosen as $u^a = v_u\delta_{a3}$, and
$\vp' = 0$. During tachyonic preheating topological defects
are formed, with a characteristic separation given by the coherence length
$k_*^{-1}$. For the symmetry breaking $\text{SU}(2)$ to $\text{U}(1)$,
these defects are monopoles.

After this phase of tachyonic preheating inflation continues with the
inflaton $\vp$ and the Hubble parameter $H = \lambda
v_s^2/(2\sqrt{3}\Mp)$. The critical value $\vp_c = v_s$ is reached at time
\begin{equation}
t_c = t_i + \frac{24\pi^2H}{m_S^4} {\vp_i}^2 \gg t'_c \ .
\end{equation}
Writing $S=(S^0,S^-)^T$ and $S_c = (S^0_c,S^+)^T$,
the scalar potential at $\vp \sim v_s$ becomes
\begin{equation}
\begin{split}
  \mathcal{V}_F = \frac{\lambda^2}{4}v_s^4
  &+  \frac{\lambda^2}{2}\vp^2 \big(|S^0|^2+|S^0_c|^2\big)
  + \frac{1}{2}\big(\lambda\vp + 2\sqrt{2}hv_u\big)^2 \big(|S^-|^2 + |S^+|^2\big)\\
   & -\lambda^2v_s^2 \big(\text{Re}[S^0_cS^0] + \text{Re}[S^+S^-]\big)
   + \ldots .
\end{split}
\end{equation}
Since $\text{SU}(2)$ is broken the mass matrices for the
$S^0_c$-$S^0$ system and the $S^+$-$S^-$ system are different. 
For $hv_u \gtrsim \lambda v_s$ the $S^+$-$S^-$ mass matrix has only positive
eigenvalues, and the fields $S^+$ and $S^-$ therefore stay at zero.
In the $S^0_c$-$S^0$ system it is convenient to introduce the linear combinations
$S_{1,2} = (\text{Re}[S^0\pm S^0_c] + i \text{Im}[S^0 \mp
S^0_c])/\sqrt{2}$. The scalar potential for the fields $S_{1,2}$, $\vp$ and $\vp'$
 reads
\begin{equation}\label{tachyon2}
  \begin{split}
    \mathcal{V}_F &= \frac{1}{4}\left(h(|S_1|^2 - |S_2|^2)
      - \sqrt{2}\lambda'\vp'v_u\right)^2
    + \frac{h^2 + \lambda^2}{4} (\text{Im}[S_1S_2])^2\\
&+ \frac{\lambda^2}{4}\left(v_s^4 + 2(\vp^2 - v_s^2)|S_1|^2 + 2(\vp^2 + v_s^2)|S_2|^2
+ (|S_1|^2 - |S_2|^2)^2 \right)\ .
    \end{split}
  \end{equation}
 At $\vp = v_s$, the waterfall field $S_1$ becomes tachyonic whereas the
 mass squared of $S_2$ remains positive. This leads to a second phase of tachyonic
 preheating where a $\text{U}(1)$ symmetry is broken and cosmic
 strings are formed. Within a few Hubble times inflaton and waterfall
 field reach their vacuum values $S_1 = v_s$ and $\vp = 0$. The vev of
 $S_1$ generates a tadpole for $\vp'$, which is therefore shifted from
 $\vp'=0$ to $\vp' = hv_s^2/(\sqrt{2}\lambda' v_u)$.

 The total number of $e$-folds generated in the second phase of inflation is
 \begin{equation}
   N(\vp(t_c')) = \int_{t(\vp_c)}^{t_c'} Hdt
   = \frac{4\pi^2}{\lambda^2\Mp^2}(\vp(t_c')^2 - v_s^2) \simeq
   \frac{4\pi^2}{\lambda^2\Mp^2}\vp_i^2\ ,
 \end{equation}
where we have used that inflation ends at the critical field value
$\vp_c = v_s$, and the second equation holds for the considered
parameters. The field value corresponding to the last 50 $e$-folds
is denoted as $\vp_*$, i.e., $N(\vp_*) \equiv N_* = 50$.
The reheating process after inflation has been studied in
detail in \cite{Buchmuller:2012wn,Buchmuller:2013lra}. Successful
leptogenesis restricts the reheating temperature to $10^8~\text{GeV}
\lesssim T_\text{rh} \lesssim 10^{10}~\text{GeV}$. Since the reheating
process is triggered by Higgs decays, this constrains the Higgs mass,
and therefore the Higgs coupling to $\lambda \sim 10^{-3}$ within about
one order of magnitude. For Higgs couplings in this range, the inflationary
observables, given in terms of the slow-roll parameters
\begin{equation}
\epsilon = \frac{M^2_{\text{P}}}{2}\left(\frac{\partial_\vp V}{V}\right)^2 \ ,\quad
\eta = M^2_{\text{P}} \frac{\partial^2_\vp V}{V} \ ,
  \end{equation}
  satifsfy the simple relations,
  \begin{equation}\label{infobs}
    \begin{split}
  A_s &= \left.\frac{H^2}{8\pi^2\epsilon
      M^2_{\text{P}}}\right|_{\vp_*} \simeq \frac{1}{3}
  \left(\frac{v_s}{\Mp}\right)^4 N_* \ , \\
n_s &\simeq 1 + 2\eta|_{\vp_*} \simeq 1 - \frac{1}{N_*} \simeq 0.98\ ,
\end{split}
\end{equation}
where $A_s$ and $n_s$ are the amplitude of scalar fluctuations and the
scalar spectral index, respectively. With $A_s = (2.099\pm
0.029)\times 10^{-9}$
\cite{Akrami:2018odb}, one obtains from Eq.~\eqref{infobs} a
symmetry breaking vev of order the GUT scale, $v \approx 5\times 10^{15}~\text{GeV}$.
The tensor-to-scalar ratio turns out to be very small,
$r = A_t/A_s = 16\epsilon|_{\vp_*} \simeq \lambda^2/(2\pi^2N_*)
\lesssim 10^{-7}$. The predicted scalar spectral index $n_s$ is
consistent at $3\sigma$ CL with the observed value
$n_s = 0.9649 \pm 0.0042$ \cite{Akrami:2018odb}. The agreement with
observation can be improved by taking the supergravity effect of a
constant $P_0$ in the superpotential into account, which is related to the
gravitino mass as $P_0 = \alpha m_{3/2} \Mp$, where $\alpha \sim1$
encodes the details of supersymmetry breaking. Remarkably, the
predicted scalar spectral index turns out to be consistent with observation for gravitino
masses in the range $10~\text{TeV} \lesssim m_{3/2} \lesssim
10~\text{PeV}$. Variation of the gravitino mass in this range is
connected to a variation of the Higgs coupling $\lambda$ and the
reheating temperature by about two orders of magnitude.
A detailed discussion can be found in 
\cite{Buchmuller:2014epa,Buchmuller:2019gfy}.

For generic initial values $\vp_i \gg v_s$,
the total number of $e$-folds during the second phase of inflation
is much larger than $50$, and the original
density of monopoles is completely diluted. The strings produced in
the second phase of tachyonic preheating are metastable, i.e., long
string loops tunnel to configurations of equal energy, consisting of
string segments that connect monopole-antimonopole pairs
\cite{Preskill:1992ck}. The geometry of these configurations is
reviewed in \cite{Kibble:2015twa}. Consider the monopole at one end
of a string segment. At large distances away from the center of the
monopole the triplet field has the familiar hedgehog structure \eqref{hedgehog},
so that in polar coordinates $(r,\alpha,\beta)$,
\begin{equation}
  \tilde{U} \propto \frac{x^a}{r} \tau^a = \left(\begin{array}{c c}
        \cos{\alpha} & \sin{\alpha} e^{-i\beta}\\
        \sin{\alpha} e^{i\beta} & - \cos{\alpha} \end{array}\right) \ .
  \end{equation}
  To minimize the energy density of the configuration, $U$, $S$ and
  $S_c$ have to be ``aligned'' around the sphere. From the scalar potential
  \eqref{Fpotential} one reads off that $S$ and $S_c$ have to be
  eigenvectors of $\tilde{U}$ with postive eigenvalue in all directions, which
  corresponds to \cite{Nambu:1977ag}
    \begin{equation}
    S_c = S \propto \left(\begin{array}{c} \cos{\frac{\alpha}{2}} \\
                      \sin{\frac{\alpha}{2}} e^{i\beta} \end{array}\right) \ .
                \end{equation}
Because of the phase factor, these eigenvectors are singular at the
south pole $\alpha = \pi$.
Here a string must be attached, around which the phases of $S$ and
$S_c$ change by $2\pi$.

\subsection{Formation of dumbbells}

Consider now the opposite case $m_S \gg m_U$. Now $\vp$ reaches
its critical value $\vp_c \sim v_s$ first, at a time
\begin{equation}
t_c = t_i + \frac{24\pi^2H_i}{m_S^4} {\vp_i}^2 \ ,
\end{equation}
when $\vp'$ is still much larger than $v_u$. For $\vp \sim \vp_c$,
the scalar potential reads
\begin{equation}
\begin{split}
  \mathcal{V}_F &\simeq \frac{{\lambda'}^2}{4} {v_u}^4 
  + \frac{{\lambda'}^2}{2} {\vp'}^2 U^\dagger U 
 + \frac{3 m_U^4}{128\pi^2} \ln{\left(\frac{{\vp'}^2}{v_u^2}\right)}
\\
  &+ \frac{\lambda^2}{4} \left(v_s^4 
  + 2 \hat{\vp}^2 \big(S^\dagger S +S_{c}^\dagger S_c\big)
   - 4v_s^2 \text{Re}[S^T_{c}S] + 4 |S^T_{c}S|^2\right) + \ldots \ ,
\end{split}
\end{equation}
where $\hat{\vp} = \vp + \sqrt{2}hv_u/\lambda$. The potential
is invariant under $\text{SU}(2)\times \text{U}(1)$.
At a critical value $\vp_c$, the waterfall fields $S$ and $S_c$ become
tachyonic along the D-flat directions. Their vev's can be chosen to
be the upper components of $S$ and $S_c$. Switching again from $S^0$
and $S^0_c$ to $S_{1,2}$, the scalar potential becomes
\begin{equation}\label{tachyon3}
\begin{split}
  \mathcal{V}_F \simeq &\frac{{\lambda'}^2}{4} {v_u}^4
  + \frac{{\lambda'}^2}{2} {\vp'}^2 U^\dagger U 
  + \frac{3 m_U^4}{128\pi^2} \ln{\left(\frac{{\vp'}^2}{v_u^2}\right)}
  + \frac{\lambda^2}{4}\big(v_s^4 + 2(\hat{\vp}^2 - v_s^2)|S_1|^2 \\
    &+ 2(\hat{\vp}^2 + v_s^2)|S_2|^2
+ (|S_1|^2 - |S_2|^2)^2   + (\text{Im}[S_1S_2])^2 \big) + \ldots\ .
\end{split}
\end{equation}
At the critical value $\vp_c= v_s - \sqrt{2}hv_u/\lambda$, the field $S_1$
becomes tachyonic and in a few Hubble times the system reaches the
vacuum configuration $S_1= v_s$, $\vp = -\sqrt{2}hv_u/\lambda$.
The field $S_2$ stays at zero.

The symmetry is now broken from $\text{SU}(2)\times
\text{U}(1)/\mathbb{Z}_2$ to $\text{U}(1)$ and the vacuum manifold is
$\text{U}(2)/\text{U}(1) = S^3$. As discussed in the previous section,
there are no topologically stable strings, but instead dumbbells or 
X-strings, analogous to the Z-strings of the Standard Model. These
defects are unstable\footnote{For a numerical simulation of dumbbell
  production during a phase transition, see \cite{Copeland:1987ht}.}.

After tachyonic preheating inflation continues with $\vp'$ as inflaton and
the critical value $\vp'_c$ is reached at time
\begin{equation}
t'_c \simeq t_i + \frac{32\pi^2H'}{m_U^4} {\vp'_i}^2  \ ,
\end{equation}
where now the Hubble parameter is $H' = \lambda' v_u^2/(2\sqrt{3}\Mp)$.

With $S$ and $S_c$ settled at their vacuum values, and splitting again
the waterfall fields $U^a$ into real and imaginary parts, $U^a = u^a + iw^a$,
the scalar potential for $u^a$ and $\vp'$ reads
\begin{equation}
  \mathcal{V}_F = \frac{h^2}{4} v_s^4 + \frac{3 m_U^4}{128\pi^2}
  \ln{\left(\frac{{\vp'}^2}{v_u^2}\right)} - \frac{h\lambda'}{\sqrt{2}} v_s^2 \vp' u^3
+ \frac{{\lambda'}^2}{2}{\vp'}^2 u^a u^a + \ldots   
\end{equation}
Contrary to the potentials \eqref{tachyon1}, \eqref{tachyon2} and
\eqref{tachyon3}, now a term linear in the waterfall field $U$ appears.
This reflects the fact that after the condensation of $S$ and $S_c$
the symmetry $\text{U}(2)$ is already broken to the final $\text{U}(1)$ subgroup,
and hence also $u^3$ gets a small vev which for large values of $\vp'$ reads
\begin{equation}
u^3 \simeq \frac{\sqrt{2}}{h\lambda'v_s^2} \frac{1}{\vp'}   
\left(\frac{h^2}{4} v_s^4 + \frac{3 m_U^4}{128\pi^2}
  \ln{\left(\frac{{\vp'}^2}{{v_u}^2}\right)}\right) + \dots
\end{equation}
As $\vp'$ decreases, $u^3$ increases until eventually all fields are
of the order of $v_u$, $v_s$. The scalar potential for the fields
$u^a$, $w^a$, $\vp$ and $\vp'$ is given by ($i = 1,2$, $a = 1,..,3$)
\begin{equation}
  \begin{split}
  \mathcal{V}_F = &\Big(\frac{h}{2} v_s^2  - \frac{\lambda'}{\sqrt{2}}
    \vp' u^3\Big)^2
+ \frac{{\lambda'}^2}{4} \Big(\big(v_u^2-u^au^a+w^aw^a\big)^2
  + 4\big(u^a w^a\big)^2\Big) \\
&+\frac{{{\lambda'}^2}}{2} {\vp'}^2 \big(u^iu^i + w^aw^a\big) 
+ v_s^2\Big(\big(hu^3 - \frac{\lambda}{\sqrt{2}}\vp - hv_u\big)^2
    + h^2\big(u^iu^i + w^aw^a\big) \Big) \ .
  \end{split}
\end{equation}
Backreaction between waterfall fields and inflaton fields leads to
oscillations, and eventually the fields settle in the supersymmetric
vacuum $u^3 = v_u$, $u^i = w^a = 0$, $\vp = 0$ and $\vp' =
hv_s^2/(\sqrt{2}v_u)$. Hence, there is no second phase of tachyonic 
preheating and there are no topological defects that survive inflation.
Therefore, there is no SGWB produced by a string network and also the
reheating mechanism described in the previous section is lost.

\section{Conclusions}

We have studied symmetry breaking and topological defects in a
supersymmetric model with gauge group $\text{U}(2) = \text{SU}(2)_R
\times \text{U}(1)_{B-L}/\mathbb{Z}_2$, which is the simplest non-Abelian extension
of a previously considered model with local $\text{U}(1)_{B-L}$
symmetry. The model has a triplet $U$ and two doublets, $S$ and $S_c$
of chiral superfields. To lift the vacuum degeneracy of the D-term
potential two gauge singlet chiral superfields are needed, which can
naturally play the role of inflatons.

The model has various topological defects. The breaking of
$\text{SU}(2)$ to $\text{U}(1)$ by a vev of $U$ leads to monopoles,
and the subsequent breaking of $\text{U}(1) \times \text{U}(1)$ to
$\text{U}(1)$ by vev's of $S$ and $S_c$ leads to metastable strings. They are 
classically stable but decay by quantum tunneling. The breaking of
$\text{U}(2)$ to $\text{U}(1)$ by vev's of $S$ and $S_c$ only leads to
unstable dumbbells or X-strings, analogous to Z-strings in the Standard Model.

Which sequence of symmetry breakings is realized depends for thermal
phase transitions on the ratio of Higgs masses $m^2_S/m^2_U$. The same
is true for hybrid inflation. Starting at large initial values of the
two inflaton fields, the speed of their slow-roll motion is
proportinal to $m_S^4$ and $m_U^4$, respectively. For $m_U \gg m_S$,
first $U$ condenses and monopoles are formed in tachyonic preheating.
In the subsequent phase of inflation the monopoles are diluted and 
in the second phase of tachyonic preheating metastable strings are
produced. Alternatively, for $m_S^2 \gg m_U^2$, first $S$ and $S_c$
condense and unstable dumbbells are formed. In the second
phase of inflation, the expectation value of $U$ smoothly evolves from
zero to its vacuum value and there is no second phase of tachyonic
preheating.

Only in the case $m_U \gg m_S$ a metastable string network is formed
which can generate a stochastic gravitational wave background. If the
reheating process is required to include thermal or non-thermal
leptogenesis, the reheating temperature has to be larger than about
$10^8~\text{GeV}$. This implies a lower bound on the Higgs mass $m_S$
and the Higgs coupling $\lambda$, such that the string tension is
predicted in a narrow range, $G\mu = (1.0 - 5.6) \times
10^{-7}$. This leads to the prediction of a SGWB with amplitude
$h^2\Omega_{\text{GW}} \sim 10^{-8}$ in the LIGO-Virgo frequency band.
The spectrum of the SGWB depends exponentially on the
monopole-string-tension ratio $\sqrt{\kappa}$. For $\sqrt{\kappa}
\simeq 8$ the evidence for a stochastic process recently reported
by the NANOGrav collaboration can be interpreted as SGWB from a
metastable cosmic string network. However, already for $\sqrt{\kappa}
\simeq 6$ the SGWB amplitude is cut off above the LIGO-Virgo frequency
band. In the considered model, for approximately equal symmetry breaking scales,
the value $\sqrt{\kappa} \simeq 8$ is obtained from geometrical
factors and the value of the gauge coupling at the GUT scale.\\




\noindent {\large \textbf{Acknowledgments}} \medskip


\noindent I thank Valerie Domcke and Kai Schmitz for
valuable discussions and comments on the manuscript.



\bibliographystyle{JHEP}
\bibliography{draft}

\providecommand{\href}[2]{#2}\begingroup\raggedright\begin{thebibliography}{10}

\bibitem{Kibble:1976sj}
T.~Kibble, {\it {Topology of Cosmic Domains and Strings}},  {\em J. Phys. A}
  {\bf 9} (1976) 1387--1398.

\bibitem{Hindmarsh:2011qj}
M.~Hindmarsh, {\it {Signals of Inflationary Models with Cosmic Strings}},  {\em
  Prog. Theor. Phys. Suppl.} {\bf 190} (2011) 197--228,
  [\href{http://arxiv.org/abs/1106.0391}{{\tt arXiv:1106.0391}}].

\bibitem{Auclair:2019wcv}
P.~Auclair et~al., {\it {Probing the gravitational wave background from cosmic
  strings with LISA}},  {\em JCAP} {\bf 2004} (2020) 034,
  [\href{http://arxiv.org/abs/1909.00819}{{\tt arXiv:1909.00819}}].

\bibitem{Nielsen:1973cs}
H.~B. Nielsen and P.~Olesen, {\it {Vortex Line Models for Dual Strings}},  {\em
  Nucl. Phys.} {\bf B61} (1973) 45--61.

\bibitem{tHooft:1974kcl}
G.~'t~Hooft, {\it {Magnetic Monopoles in Unified Gauge Theories}},  {\em Nucl.
  Phys.} {\bf B79} (1974) 276--284.

\bibitem{Polyakov:1974ek}
A.~M. Polyakov, {\it {Particle Spectrum in the Quantum Field Theory}},  {\em
  JETP Lett.} {\bf 20} (1974) 194--195. [Pisma Zh. Eksp. Teor.
  Fiz.20,430(1974)].

\bibitem{Nambu:1977ag}
Y.~Nambu, {\it {String-Like Configurations in the Weinberg-Salam Theory}},
  {\em Nucl. Phys.} {\bf B130} (1977) 505.

\bibitem{Kibble:1982ae}
T.~W.~B. Kibble, G.~Lazarides, and Q.~Shafi, {\it {Strings in SO(10)}},  {\em
  Phys. Lett.} {\bf 113B} (1982) 237--239.

\bibitem{Copeland:1994vg}
E.~J. Copeland, A.~R. Liddle, D.~H. Lyth, E.~D. Stewart, and D.~Wands, {\it
  {False vacuum inflation with Einstein gravity}},  {\em Phys. Rev.} {\bf D49}
  (1994) 6410--6433, [\href{http://arxiv.org/abs/astro-ph/9401011}{{\tt
  astro-ph/9401011}}].

\bibitem{Dvali:1994ms}
G.~R. Dvali, Q.~Shafi, and R.~K. Schaefer, {\it {Large scale structure and
  supersymmetric inflation without fine tuning}},  {\em Phys. Rev. Lett.} {\bf
  73} (1994) 1886--1889, [\href{http://arxiv.org/abs/hep-ph/9406319}{{\tt
  hep-ph/9406319}}].

\bibitem{Jeannerot:2003qv}
R.~Jeannerot, J.~Rocher, and M.~Sakellariadou, {\it {How generic is cosmic
  string formation in SUSY GUTs}},  {\em Phys. Rev. D} {\bf 68} (2003) 103514,
  [\href{http://arxiv.org/abs/hep-ph/0308134}{{\tt hep-ph/0308134}}].

\bibitem{Minkowski:1977sc}
P.~Minkowski, {\it {$\mu \to e\gamma$ at a Rate of One Out of $10^{9}$ Muon
  Decays?}},  {\em Phys. Lett.} {\bf 67B} (1977) 421--428.

\bibitem{Yanagida:1979as}
T.~Yanagida, {\it {Horizontal gauge symmetry and masses of neutrinos}},  {\em
  Conf. Proc.} {\bf C7902131} (1979) 95--99.

\bibitem{GellMann:1980vs}
M.~Gell-Mann, P.~Ramond, and R.~Slansky, {\it {Complex Spinors and Unified
  Theories}},  {\em Conf. Proc.} {\bf C790927} (1979) 315--321,
  [\href{http://arxiv.org/abs/1306.4669}{{\tt arXiv:1306.4669}}].

\bibitem{Fukugita:1986hr}
M.~Fukugita and T.~Yanagida, {\it {Baryogenesis Without Grand Unification}},
  {\em Phys. Lett.} {\bf B174} (1986) 45--47.

\bibitem{Bodeker:2020ghk}
D.~Bodeker and W.~Buchmuller, {\it {Baryogenesis from the weak scale to the GUT
  scale}},  \href{http://arxiv.org/abs/2009.07294}{{\tt arXiv:2009.07294}}.

\bibitem{Buchmuller:2012wn}
W.~Buchmuller, V.~Domcke, and K.~Schmitz, {\it {Spontaneous B-L Breaking as the
  Origin of the Hot Early Universe}},  {\em Nucl. Phys.} {\bf B862} (2012)
  587--632, [\href{http://arxiv.org/abs/1202.6679}{{\tt arXiv:1202.6679}}].

\bibitem{Buchmuller:2014epa}
W.~Buchmuller, V.~Domcke, K.~Kamada, and K.~Schmitz, {\it {Hybrid Inflation in
  the Complex Plane}},  {\em JCAP} {\bf 1407} (2014) 054,
  [\href{http://arxiv.org/abs/1404.1832}{{\tt arXiv:1404.1832}}].

\bibitem{Buchmuller:2013lra}
W.~Buchmuller, V.~Domcke, K.~Kamada, and K.~Schmitz, {\it {The Gravitational
  Wave Spectrum from Cosmological $B-L$ Breaking}},  {\em JCAP} {\bf 1310}
  (2013) 003, [\href{http://arxiv.org/abs/1305.3392}{{\tt arXiv:1305.3392}}].

\bibitem{Dror:2019syi}
J.~A. Dror, T.~Hiramatsu, K.~Kohri, H.~Murayama, and G.~White, {\it {Testing
  the Seesaw Mechanism and Leptogenesis with Gravitational Waves}},  {\em Phys.
  Rev. Lett.} {\bf 124} (2020), no.~4 041804,
  [\href{http://arxiv.org/abs/1908.03227}{{\tt arXiv:1908.03227}}].

\bibitem{Arzoumanian:2018saf}
{\bf NANOGRAV} Collaboration, Z.~Arzoumanian et~al., {\it {The NANOGrav 11-year
  Data Set: Pulsar-timing Constraints On The Stochastic Gravitational-wave
  Background}},  {\em Astrophys. J.} {\bf 859} (2018), no.~1 47,
  [\href{http://arxiv.org/abs/1801.02617}{{\tt arXiv:1801.02617}}].

\bibitem{Kerr:2020qdo}
M.~Kerr et~al., {\it {The Parkes Pulsar Timing Array Project: Second data
  release}},  {\em Publ. Astron. Soc. Austral.} {\bf 37} (2020) e020,
  [\href{http://arxiv.org/abs/2003.09780}{{\tt arXiv:2003.09780}}].

\bibitem{Shannon:2015ect}
R.~M. Shannon et~al., {\it {Gravitational waves from binary supermassive black
  holes missing in pulsar observations}},  {\em Science} {\bf 349} (2015),
  no.~6255 1522--1525, [\href{http://arxiv.org/abs/1509.07320}{{\tt
  arXiv:1509.07320}}].

\bibitem{Buchmuller:2019gfy}
W.~Buchmuller, V.~Domcke, H.~Murayama, and K.~Schmitz, {\it {Probing the scale
  of grand unification with gravitational waves}},  {\em Phys. Lett.} {\bf B}
  (2020) 135764, [\href{http://arxiv.org/abs/1912.03695}{{\tt
  arXiv:1912.03695}}].

\bibitem{Vilenkin:1982hm}
A.~Vilenkin, {\it {Cosmological evolution of monopoles connected by strings}},
  {\em Nucl. Phys.} {\bf B196} (1982) 240--258.

\bibitem{Preskill:1992ck}
J.~Preskill and A.~Vilenkin, {\it {Decay of metastable topological defects}},
  {\em Phys. Rev.} {\bf D47} (1993) 2324--2342,
  [\href{http://arxiv.org/abs/hep-ph/9209210}{{\tt hep-ph/9209210}}].

\bibitem{Leblond:2009fq}
L.~Leblond, B.~Shlaer, and X.~Siemens, {\it {Gravitational Waves from Broken
  Cosmic Strings: The Bursts and the Beads}},  {\em Phys. Rev.} {\bf D79}
  (2009) 123519, [\href{http://arxiv.org/abs/0903.4686}{{\tt
  arXiv:0903.4686}}].

\bibitem{Monin:2008mp}
A.~Monin and M.~B. Voloshin, {\it {The Spontaneous breaking of a metastable
  string}},  {\em Phys. Rev.} {\bf D78} (2008) 065048,
  [\href{http://arxiv.org/abs/0808.1693}{{\tt arXiv:0808.1693}}].

\bibitem{Blanco-Pillado:2013qja}
J.~J. Blanco-Pillado, K.~D. Olum, and B.~Shlaer, {\it {The number of cosmic
  string loops}},  {\em Phys. Rev.} {\bf D89} (2014), no.~2 023512,
  [\href{http://arxiv.org/abs/1309.6637}{{\tt arXiv:1309.6637}}].

\bibitem{Abbott:2021ksc}
{\bf LIGO Scientific, Virgo, KAGRA} Collaboration, R.~Abbott et~al., {\it
  {Constraints on cosmic strings using data from the third Advanced LIGO-Virgo
  observing run}},  \href{http://arxiv.org/abs/2101.12248}{{\tt
  arXiv:2101.12248}}.

\bibitem{Arzoumanian:2020vkk}
{\bf NANOGrav} Collaboration, Z.~Arzoumanian et~al., {\it {The NANOGrav 12.5~yr
  Data Set: Search for an Isotropic Stochastic Gravitational-wave Background}},
   {\em Astrophys. J. Lett.} {\bf 905} (2020), no.~2 L34,
  [\href{http://arxiv.org/abs/2009.04496}{{\tt arXiv:2009.04496}}].

\bibitem{Ellis:2020ena}
J.~Ellis and M.~Lewicki, {\it {Cosmic String Interpretation of NANOGrav Pulsar
  Timing Data}},  {\em Phys. Rev. Lett.} {\bf 126} (2021) 041304,
  [\href{http://arxiv.org/abs/2009.06555}{{\tt arXiv:2009.06555}}].

\bibitem{Blasi:2020mfx}
S.~Blasi, V.~Brdar, and K.~Schmitz, {\it {Has NANOGrav found first evidence for
  cosmic strings?}},  {\em Phys. Rev. Lett.} {\bf 126} (2021) 041305,
  [\href{http://arxiv.org/abs/2009.06607}{{\tt arXiv:2009.06607}}].

\bibitem{Buchmuller:2020lbh}
W.~Buchmuller, V.~Domcke, and K.~Schmitz, {\it {From NANOGrav to LIGO with
  metastable cosmic strings}},  {\em Phys. Lett.} {\bf B811} (2020) 135914,
  [\href{http://arxiv.org/abs/2009.10649}{{\tt arXiv:2009.10649}}].

\bibitem{Blanco-Pillado:2021ygr}
J.~J. Blanco-Pillado, K.~D. Olum, and J.~M. Wachter, {\it {Comparison of cosmic
  string and superstring models to NANOGrav 12.5-year results}},
  \href{http://arxiv.org/abs/2102.08194}{{\tt arXiv:2102.08194}}.

\bibitem{Chigusa:2020rks}
S.~Chigusa, Y.~Nakai, and J.~Zheng, {\it {Implications of Gravitational Waves
  for Supersymmetric Grand Unification}},
  \href{http://arxiv.org/abs/2011.04090}{{\tt arXiv:2011.04090}}.

\bibitem{Chakrabortty:2020otp}
J.~Chakrabortty, G.~Lazarides, R.~Maji, and Q.~Shafi, {\it {Primordial
  Monopoles and Strings, Inflation, and Gravity Waves}},
  \href{http://arxiv.org/abs/2011.01838}{{\tt arXiv:2011.01838}}.

\bibitem{Samanta:2020cdk}
R.~Samanta and S.~Datta, {\it {Gravitational wave complementarity and impact of
  NANOGrav data on gravitational leptogenesis: cosmic strings}},
  \href{http://arxiv.org/abs/2009.13452}{{\tt arXiv:2009.13452}}.

\bibitem{King:2020hyd}
S.~F. King, S.~Pascoli, J.~Turner, and Y.-L. Zhou, {\it {Gravitational Waves
  and Proton Decay: Complementary Windows into Grand Unified Theories}},  {\em
  Phys. Rev. Lett.} {\bf 126} (2021), no.~2 021802,
  [\href{http://arxiv.org/abs/2005.13549}{{\tt arXiv:2005.13549}}].

\bibitem{Blasi:2020wpy}
S.~Blasi, V.~Brdar, and K.~Schmitz, {\it {Fingerprint of low-scale leptogenesis
  in the primordial gravitational-wave spectrum}},  {\em Phys. Rev. Res.} {\bf
  2} (2020), no.~4 043321, [\href{http://arxiv.org/abs/2004.02889}{{\tt
  arXiv:2004.02889}}].

\bibitem{Cui:2017ufi}
Y.~Cui, M.~Lewicki, D.~E. Morrissey, and J.~D. Wells, {\it {Cosmic Archaeology
  with Gravitational Waves from Cosmic Strings}},  {\em Phys. Rev.} {\bf D97}
  (2018), no.~12 123505, [\href{http://arxiv.org/abs/1711.03104}{{\tt
  arXiv:1711.03104}}].

\bibitem{Cui:2018rwi}
Y.~Cui, M.~Lewicki, D.~E. Morrissey, and J.~D. Wells, {\it {Probing the pre-BBN
  universe with gravitational waves from cosmic strings}},  {\em JHEP} {\bf 01}
  (2019) 081, [\href{http://arxiv.org/abs/1808.08968}{{\tt arXiv:1808.08968}}].

\bibitem{Gouttenoire:2019kij}
Y.~Gouttenoire, G.~Servant, and P.~Simakachorn, {\it {Beyond the Standard
  Models with Cosmic Strings}},  {\em JCAP} {\bf 07} (2020) 032,
  [\href{http://arxiv.org/abs/1912.02569}{{\tt arXiv:1912.02569}}].

\bibitem{Gouttenoire:2019rtn}
Y.~Gouttenoire, G.~Servant, and P.~Simakachorn, {\it {BSM with Cosmic Strings:
  Heavy, up to EeV mass, Unstable Particles}},  {\em JCAP} {\bf 07} (2020) 016,
  [\href{http://arxiv.org/abs/1912.03245}{{\tt arXiv:1912.03245}}].

\bibitem{Felder:2000hj}
G.~N. Felder, J.~Garcia-Bellido, P.~B. Greene, L.~Kofman, A.~D. Linde, and
  I.~Tkachev, {\it {Dynamics of symmetry breaking and tachyonic preheating}},
  {\em Phys. Rev. Lett.} {\bf 87} (2001) 011601,
  [\href{http://arxiv.org/abs/hep-ph/0012142}{{\tt hep-ph/0012142}}].

\bibitem{Wess:1992cp}
J.~Wess and J.~Bagger, {\em {Supersymmetry and supergravity}}.
\newblock Princeton University Press, Princeton, NJ, USA, 1992.

\bibitem{Seiberg:1994rs}
N.~Seiberg and E.~Witten, {\it {Electric - magnetic duality, monopole
  condensation, and confinement in N=2 supersymmetric Yang-Mills theory}},
  {\em Nucl. Phys.} {\bf B426} (1994) 19--52,
  [\href{http://arxiv.org/abs/hep-th/9407087}{{\tt hep-th/9407087}}]. [Erratum:
  Nucl. Phys.B430,485(1994)].

\bibitem{Bogomolny:1975de}
E.~B. Bogomolny, {\it {Stability of Classical Solutions}},  {\em Sov. J. Nucl.
  Phys.} {\bf 24} (1976) 449. [Yad. Fiz.24,861(1976)].

\bibitem{Prasad:1975kr}
M.~K. Prasad and C.~M. Sommerfield, {\it {An Exact Classical Solution for the
  't Hooft Monopole and the Julia-Zee Dyon}},  {\em Phys. Rev. Lett.} {\bf 35}
  (1975) 760--762.

\bibitem{Copeland:1987ht}
E.~J. Copeland, D.~Haws, T.~W.~B. Kibble, D.~Mitchell, and N.~Turok, {\it
  {Monopoles Connected by Strings and the Monopole Problem}},  {\em Nucl.
  Phys.} {\bf B298} (1988) 445--457.

\bibitem{Kephart:1995cg}
T.~W. Kephart and T.~Vachaspati, {\it {Topological incarnations of electroweak
  defects}},  {\em Phys. Lett.} {\bf B388} (1996) 481--486,
  [\href{http://arxiv.org/abs/hep-ph/9503355}{{\tt hep-ph/9503355}}].

\bibitem{Achucarro:1999it}
A.~Achucarro and T.~Vachaspati, {\it {Semilocal and electroweak strings}},
  {\em Phys. Rept.} {\bf 327} (2000) 347--426,
  [\href{http://arxiv.org/abs/hep-ph/9904229}{{\tt hep-ph/9904229}}]. [Phys.
  Rept.327,427(2000)].

\bibitem{Kibble:2015twa}
T.~W.~B. Kibble and T.~Vachaspati, {\it {Monopoles on strings}},  {\em J.
  Phys.} {\bf G42} (2015), no.~9 094002,
  [\href{http://arxiv.org/abs/1506.02022}{{\tt arXiv:1506.02022}}].

\bibitem{Aulakh:1998nn}
C.~S. Aulakh, A.~Melfo, and G.~Senjanovic, {\it {Minimal supersymmetric
  left-right model}},  {\em Phys. Rev.} {\bf D57} (1998) 4174--4178,
  [\href{http://arxiv.org/abs/hep-ph/9707256}{{\tt hep-ph/9707256}}].

\bibitem{Babu:2008ep}
K.~S. Babu and R.~N. Mohapatra, {\it {Minimal Supersymmetric Left-Right
  Model}},  {\em Phys. Lett.} {\bf B668} (2008) 404--409,
  [\href{http://arxiv.org/abs/0807.0481}{{\tt arXiv:0807.0481}}].

\bibitem{Vachaspati:1992fi}
T.~Vachaspati, {\it {Vortex solutions in the Weinberg-Salam model}},  {\em
  Phys. Rev. Lett.} {\bf 68} (1992) 1977--1980. [Erratum: Phys. Rev.
  Lett.69,216(1992)].

\bibitem{Asaka:2001ez}
T.~Asaka, W.~Buchmuller, and L.~Covi, {\it {False vacuum decay after
  inflation}},  {\em Phys. Lett.} {\bf B510} (2001) 271--276,
  [\href{http://arxiv.org/abs/hep-ph/0104037}{{\tt hep-ph/0104037}}].

\bibitem{Akrami:2018odb}
{\bf Planck} Collaboration, Y.~Akrami et~al., {\it {Planck 2018 results. X.
  Constraints on inflation}},  {\em Astron. Astrophys.} {\bf 641} (2020) A10,
  [\href{http://arxiv.org/abs/1807.06211}{{\tt arXiv:1807.06211}}].

\end{thebibliography}\endgroup

\end{document}